\documentstyle[twocolumn,prl,aps,psfig]{revtex}

\begin{document}

\twocolumn[
\hsize\textwidth\columnwidth\hsize\csname@twocolumnfalse\endcsname
%
%\draft
\begin{flushright}
{\small SNS'97 (Spectroscopies in Novel Superconductors), Sept. 14-18, Cape 
Cod;\\
Proceedings to appear in J. Phys. Chem. Solids}\\
\end{flushright}

\title
{EVOLUTION OF MAGNETIC AND SUPERCONDUCTING FLUCTUATIONS WITH DOPING OF 
HIGH-$T_{c}$ SUPERCONDUCTORS \\
(An electronic Raman scattering study)
}
\author{
G.~BLUMBERG,\hspace*{-1mm}$^{1,2,3,\dag}$ M.V.~KLEIN,\hspace*{-1mm}$^{1,2}$  
K.~KADOWAKI,\hspace*{-1mm}$^{4}$ C.~KENDZIORA,\hspace*{-1mm}$^{5}$ 
P.~GUPTASARMA,\hspace*{-1mm}$^{1,6}$ and D.~HINKS$^{1,6}$  \\  
$^{1}$NSF Science and Technology Center for Superconductivity \\
$^{2}$Department of Physics, 
University of Illinois at Urbana-Champaign, Urbana, IL 61801-3080\\
$^{3}$Institute of Chemical Physics and Biophysics, R\"avala 10, 
Tallinn EE0001, Estonia\\ 
$^{4}$Institute of Materials Science, University of Tsukuba, Tsukuba, 
Ibaraki 305, Japan\\ 
$^{5}$Code 6653, Naval Research Laboratory, Washington, D.C. 20375\\
$^{6}$Materials Science Division, Argonne National Laboratory, 
Argonne, IL 60439
}
\maketitle

\begin{abstract}

\textbf{Abstract}--For YBa$_{2}$Cu$_{3}$O$_{6 + \delta}$ and 
Bi$_{2}$Sr$_{2}$CaCu$_{2}$O$_{8 \pm \delta}$ superconductors,    
electronic Raman scattering from  high- and low-energy 
excitations has been studied in relation to the hole doping level, 
temperature, and energy of the incident photons. 
For underdoped superconductors, it is concluded that   
short range antiferromagnetic (AF) correlations persist with hole doping 
and doped single holes are incoherent in the AF environment.  
Above the superconducting (SC) transition temperature $T_{c}$ the 
system exhibits a sharp Raman resonance of $B_{1g}$ symmetry and about 75~meV 
energy and a pseudogap for electron-hole excitations below 
75~meV, a manifestation of a partially coherent state forming from 
doped incoherent quasi-particles. 
The occupancy of the coherent state increases with cooling until phase 
ordering at $T_{c}$ produces a global SC state. 

\end{abstract}
 
\pacs{\emph{Keywords:} Raman scattering, antiferromagnetic correlations, 
superconducting order}
\bigskip
]

\section{INTRODUCTION}

The normal and superconducting (SC) properties of doped cuprate 
high-temperature superconductors are very different from those of 
conventional metals and are usually viewed as 
manifestations of strong electron-electron correlations. 
These correlations cause the antiferromagnetic (AF) state in the undoped 
cuprates. 
The SC coherence length,  
$\xi_{SC}$, is of the order of a few lattice spacings, 
considerably shorter than in conventional superconductors. 
The persistence of short-range AF correlations with doping  
has been thought to lead to an effective pairing mechanism 
and to the unconventional normal state properties \cite{Anderson87}.

The underdoped SC cuprates have low carrier density. 
The mean free path of the holes is shorter than their de Broglie 
wavelength which brings the underdoped cuprates into a class of ``bad 
metals'' \cite{EmeryKivelsonPRL95}. 
For cuprates a rough estimate of parameter $k_{F}\xi_{SC}$ ($k_{F}$ is Fermi 
wave vector) lead to values of about 3 to 20, two orders of magnitude 
smaller than for conventional Bardeen-Cooper-Schrieffer (BCS) 
superconductors. 
It was shown by Uemura {\em et. al.} \cite{Uemura} that, for cuprates, 
the SC transition occurs around the temperature at which 
the thermal de Broglie wavelength of the pairs is 2 to 6 times greater 
than the average interpair separation.  
A pair size comparable to the average interparticle 
spacing $k^{-1}_{F}$ brings cuprates to an intermediate regime 
between the BCS limit of large overlapping Cooper pairs ($k_{F}\xi_{SC} 
\gg 1$) and of Bose-Einstein (BE) condensation of composite bosons 
($k_{F}\xi_{SC} \ll 1$) consisting of tightly bound fermion pairs 
\cite{Uemura,Randeria}. 
As a consequence of the low carrier density and the short 
SC correlation length the transition to the 
SC state may not display a typical BCS behavior. 
In the underdoped cuprates, $T_{c}$ may 
be strongly suppressed with respect to the pairing temperature and   
be determined by phase fluctuations 
\cite{EmeryKivelson95,Schmalian97}. 
  
\section{PROBING MAGNETIC AND SUPERCONDUCTING FLUCTUATIONS BY 
ELECTRONIC RAMAN SCATTERING}

Electronic Raman scattering is a local high-energy probe for both 
(i) the short-range AF correlations 
and (ii) the SC order parameter in doped, SC samples through  
excitation across the SC gap. 

Two-magnon (2-M) Raman scattering directly probes short-wavelength 
magnetic fluctuations, which may exist without long-range 
AF order \cite{Lyons88,BlumbergSPIE}. 
The Raman process takes place through a photon-stimulated virtual 
charge-transfer (CT) excitation that exchanges two Cu spins. 
This process may also be described as creation of two interacting magnons. 
The CT excitation is the same one which virtually produces the spin 
superexchange constant $J$. 
In the AF environment with a correlation length $\xi_{AF}$ 
covering 2 to 3 lattice constants, the spin exchange process requires an energy 
of about $3J$. 
The magnetic Raman scattering peak position, intensity, and shape 
provide information about fluctuations in a state of 
short-range AF order \cite{BlumbergSPIE}. 

Electronic Raman scattering by charge fluctuations in metals 
arises from electron-hole (e-h) excitations near the Fermi surface (FS).  
For a normal Fermi liquid model of the cuprates, the scattering would have 
finite intensity only at very low 
frequencies.  
For strongly correlated systems, incoherent quasi-particle (QP) scattering 
leads to finite Raman intensity over a broad region of frequency 
\cite{Varma89,Shastry}, 
and the intensity can be used as a measure of the incoherent 
scattering. 
Indeed, for the cuprates in the normal state, an almost 
frequency independent Raman continuum has been observed that 
extends to at least 2~eV \cite{BlumbergSPIE,Staufer91,Blumberg2M94}.
In the SC state of optimally and overdoped cuprates, the 
low-frequency tail of the Raman continuum changes to reflect the SC effects. 
The opening of a SC gap reduces the incoherent inelastic 
scattering processes and reduces the strength of the continuum. 
Freed from the heavy damping, the QPs now show a gap in 
their spectral function. 
Thus, the electronic Raman spectrum of e-h-pair excitations 
acquires the so-called $2\Delta$-peak due to gap 
excitations, $2\Delta(\bf{k})$, where $\bf{k}$ is a wave vector on the FS. 
For optimally and overdoped cuprates, opening of an anisotropic 
superconducting gap, 
$2\Delta(\bf{k})$, in the e-h-pair excitation spectrum at 
wave vector $\bf{k}$ causes suppression 
of the low-frequency part of the continuum due to both a  
gapped density of states and to a drop in the low-energy QP scattering rate. 
For underdoped cuprates, only a relatively weak peak has been 
observed within the strong Raman continuum \cite{Slakey90}; 
its energy does not scale with sample $T_{c}$, and its origin has been unclear.

We studied both AF and SC correlations and found that 
for underdoped samples, the short-range AF  
correlations persist with hole doping. 
Furthermore, the presence of a Raman peak of $B_{1g}$ symmetry at $\sim 
75$~meV is interpreted as evidence that 
incoherent doped holes form above $T_{c}$ a long lived collective state 
with a sharp resonance of $B_{1g}$ ($d_{x^{2}-y^{2}}$) symmetry. 
The temperature dependence of this peak is evidence that this state 
gains phase coherence at $T_{c}$ and participates in the collective SC state. 
Binding of incoherent QPs into the coherent state  
reduces the low-frequency scattering rate and leads to a pseudogap in 
the spectra. 

\section{EXPERIMENTAL}

The YBa$_{2}$Cu$_{3}$O$_{6 + \delta}$ and 
Bi$_{2}$Sr$_{2}$CaCu$_{2}$O$_{8 \pm \delta}$ single crystals were grown 
and postannealed as described in Refs. 
\cite{LeeGinsberg,Veal90,Dabrowski,Ding,Kendziora,Guptasarma}. 
Raman measurements were made using systems described in Refs. 
\cite{Blumberg2M94,Kang96}. 
The polarization of the incoming and outgoing light with respect to 
the copper-oxygen plane is shown by the long straight arrows in Fig. 1. 
It gives mainly Raman spectra of $B_{1g}$ symmetry. 
For magnetic excitations, the $B_{1g}$ scattering channel couples to 
pairs of short wavelength magnons near the magnetic Brillouin zone 
boundary \cite{BlumbergSPIE}.  
For the e-h excitations near the FS, the Raman 
form-factor for $B_{1g}$ symmetry is peaked for ${\bf k}$ near the 
antinode wave vectors  
$\{{\bf k}_{an}\} = \{(0, \pm \frac{\pi}{a})$ and $(\pm \frac{\pi}{a}, 
0)\}$ \cite{Devereaux}, 
where the anisotropic SC gap magnitude is believed 
to reach its maximum value $\Delta_{max}$. 
Indeed, the $B_{1g}$ scattering geometry reveals interesting 
excitations in both magnetic and e-h excitation channels.  

\begin{figure}[t]
\centerline{
\psfig{figure=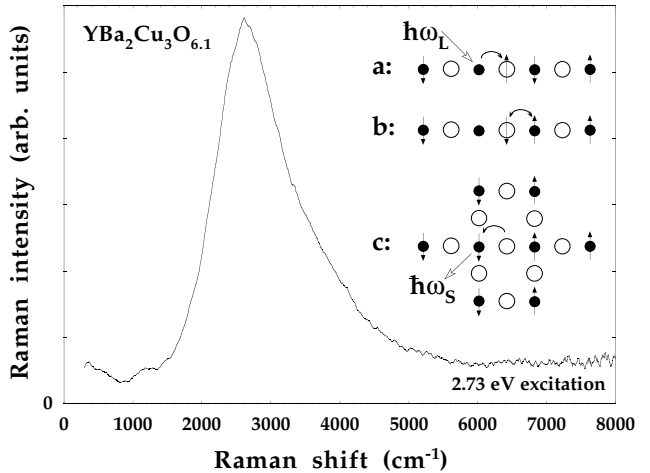,width=8.5cm,clip=}
}
\caption{
Two-magnon Raman scattering spectra from  YBa$_2$Cu$_3$O$_{6.1}$ AF insulator 
at room temperature. 
Inset illustrates the photon induced spin superexchange mechanism for 
Raman scattering and shows the polarization of the incoming laser 
light and the outgoing scattered light (long, straight arrows) with respect 
to the copper (solid circles) - oxygen (open circles) plane. 
}
\label{Fig.1}
\end{figure}

\section{ANTIFERROMAGNETIC CORRELATIONS}

Figures 1 and 2 show the high-energy part of $B_{1g}$ electronic Raman 
scattering spectra at room temperature 
as a function of hole doping. 
The typical spectrum from the AF insulator 
exhibits a strong band assigned to scattering 
by 2-Ms, that is the photon-induced superexchange of two spins 
on two nearest neighbor Cu 3d$^{9}$-orbital sites through the 
intervening 2p-oxygen orbital. 
The incoming photon can cause a charge (hole) transfer from the 
3d$^{9}$2p$^{6}$ to 3d$^{10}$2p$^{5}$ configuration (Fig.~1a). 
The probability of the virtual intermediate CT process is  
expected to be resonantly enhanced when the incoming photon energy 
$\hbar \omega_{L}$ approaches the Cu 3d -- O 2p CT  
energy $E_{CT}$ \cite{BlumbergSPIE,Pressl96}. 
In the next step (Fig.~1b) a 2p$^{5}$ and 3d$^{9}$ spin exchange occurs, 
and in the last step (Fig.~1c) the charge returns to the initial Cu 
3d$^{9}$ orbital and emits an inelasticly scattered photon  
$\hbar \omega_{S} = \hbar \omega_{L} - \hbar \omega_{2M}$. 
As a result of the superexchange, each of two exchanged spins 
sees three ferromagnetically aligned neighbors, at a cost of  
$\sim 3J$ due to the Heisenberg interaction 
energy $J \sum_{<i,j>} ({\bf S}_i \cdot {\bf S}_j - \slantfrac{1}{4})$ 
[${\bf S}_i$ is the spin on site $i$ and the summation is over near-neighbor 
Cu pairs.]. 
Thus for the AF insulators, the 2-M peak position 
yields an estimate of $J \approx 125$~meV. 

\begin{figure}[t]
\centerline{
\psfig{figure=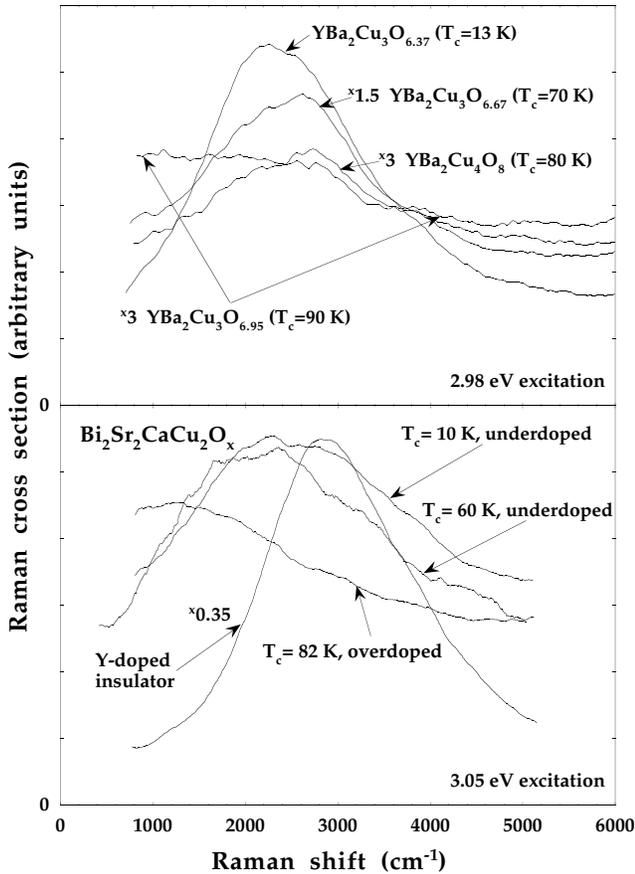,width=8.5cm,clip=}
}
\caption{ 
The $B_{1g}$ continuum and two-magnon Raman scattering spectra at 
room temperature as a function of doping for YBa$_{2}$Cu$_{3}$O$_{6 + 
\delta}$ and Bi$_{2}$Sr$_{2}$CaCu$_{2}$O$_{8 \pm \delta}$. 
}
\label{Fig.2}
\end{figure}

For doped superconductors, the spectra  (Fig.~2) exhibit a background continuum 
plus a broad peak, which, similar to the AF  
case, has been assigned to the double spin-flip excitation in 
the short-range AF environment. 
For the process of two-spin superexchange to require 
the full $3J$ energy cost, spins on six further Cu neighbor sites must 
show AF alignment. 
The doped holes are believed to form singlets with the holes that 
would otherwise form local AF order \cite{Zhang88}.
This screens the effective spin moment on the doped Cu site and might 
lead to reduction of the spin superexchange energy in the vicinity 
of holes. 
With doping, as is seen from Fig.~2, the 2-M scattering peak broadens, 
weakens, and shifts to lower frequency.  
The existence of the peak in the SC cuprates indicates 
persistence of a local short-range AF order extending a few lattice spacings. 

\begin{figure}[t]
\centerline{
\psfig{figure=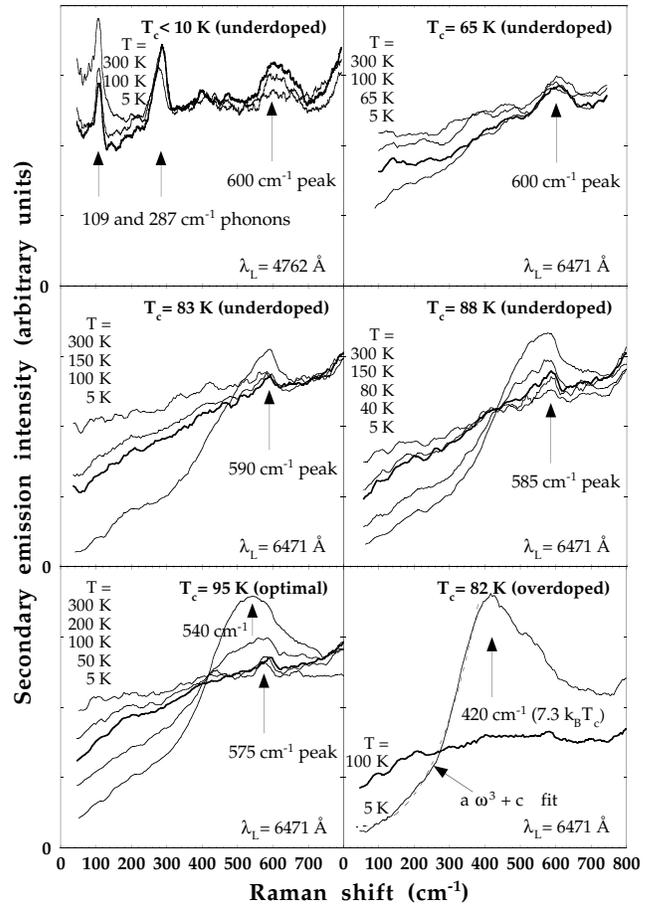,width=8.5cm,clip=}
}
\caption{ 
The low-energy portion of $B_{1g}$ Raman scattering spectra for 
Bi$_{2}$Sr$_{2}$CaCu$_{2}$O$_{8 \pm \delta}$ as a function of temperature 
and doping. 
}
\label{Fig.3}
\end{figure}

\section{INCOHERENT CONTINUUM, PSEUDOGAP AND RESONANT EXCITATION AT 
75~meV}  

Figure 3 shows the low-energy $B_{1g}$ Raman scattering spectra for 
Bi$_{2}$Sr$_{2}$CaCu$_{2}$O$_{8 \pm \delta}$ superconductors as 
function of temperature and doping.  
The most prominent feature of the spectra is the electronic continuum. 
According to the existing phenomenology for the continuum 
\cite{Varma89,Slakey91} and allowing for the symmetry of the 
$B_{1g}$ Raman form factor, we believe the $B_{1g}$ Raman continuum 
intensity, $I(\omega,T)$,  
to be proportional to the incoherent QP inverse lifetime 
$\tau_{\{{\bf k}_{an}\}}^{-1}(\omega,T)=\Gamma_{\{{\bf 
k}_{an}\}}(\omega,T) \approx 
\sqrt{(\alpha \omega)^{2}+(\beta T)^{2}}$ 
for ${\bf k}$ in the vicinity of wavevectors $\{{\bf k}_{an}\}$: 
$I(\omega,T)\propto[1+n(\omega,T)]\omega\Gamma/(\omega^{2}+\Gamma^{2})$,  
where $n(\omega,T)$ is the Bose factor, $\alpha$ and $\beta$ are 
phenomenological parameters of order unity.
The continuum extends to a few electron-volts. 
For higher temperatures it starts from very low frequencies, affirming 
strong incoherent scattering even for low lying e-h excitations. 

Cooling the underdoped samples gradually strengthens a remarkably sharp 
scattering peak at about 600~cm$^{-1}$ (75~meV).  
For underdoped YBa$_{2}$Cu$_{3}$O$_{6.6}$ a similar excitation has been 
first observed by F. Slakey {\em et al.} \cite{Slakey90}.  
The integrated (above the continuum line) intensity of the peak contains 
just a few per cent of the integrated 2-M scattering intensity.  
This peak has $B_{1g}$ symmetry and is not present for other geometries. 
The peak position shows little temperature or doping dependence. 
Below $T_{c}$ the peak undergos further intensity enhancement, turning 
into $2\Delta$-peak at higher dopings. 

The narrow width of the 600~cm$^{-1}$ peak in the Raman spectra   
($\Gamma_{peak} \lesssim$~50~cm$^{-1}$) is more than an order of 
magnitude smaller than the inverse QP lifetime 
$\Gamma_{\{{\bf k}_{an}\}}(600~cm^{-1},T)$ associated with the continuum. 
The two very different lifetimes for the 600~cm$^{-1}$ mode 
and the continuum suggest that we are essentially dealing with a two component 
system: A spectroscopically well defined, long lived, 600~cm$^{-1}$ mode on 
top of the strong incoherent background. 

\begin{figure}[t]
\centerline{
\psfig{figure=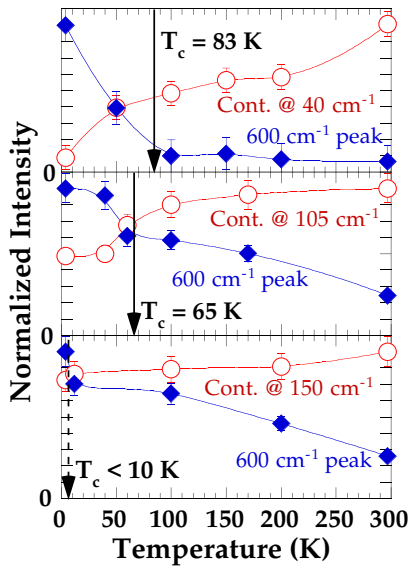,width=5.5cm,clip=}
}
\caption{ 
The temperature dependence of the normalized low-frequency continuum 
intensity (circles)  
and a normalized integrated above the continuum 600~cm$^{-1}$ peak 
intensity (diamonds) for the $T_{c} < 10$~K, 65 and 83~K underdoped samples. 
Vertical arrows denote $T_{c}$.  
}
\label{Fig.4}
\end{figure}

In Fig. 4 we plot the temperature dependence of the integrated above 
the continuum line intensity of the 600~cm$^{-1}$ mode 
for the three underdoped samples with $T_{c} < 10$~K, 65 and 83~K.  
We also plot the temperature dependence of the low-frequency continuum 
intensity.  
The integrated intensity of the 600~cm$^{-1}$ mode smoothly increases 
with cooling until the temperature reaches $T_{c}$ where the intensity 
shows a sudden enhancement.  
With cooling, the low-frequency portion of the continuum simultaneously 
shows an intensity reduction. 
The intensity reduction is an indication of a drop in the low-frequency 
($\omega < 600$~cm$^{-1}$) inverse 
lifetime $\tau_{\{{\bf k}_{an}\}}^{-1}(\omega,T)$.  
The suppression of the low-frequency spectral weight in Raman spectra 
is similar to the pseudogap phenomena that has been observed in 
spin-excitation spectra  
\cite{Warren89,Rossat-Mingnod91}, as well as in optical 
\cite{Puchkov96}, ARPES \cite{Ding,Loeser96} and 
tunneling \cite{Fischer97} studies.

\section{SUPERCONDUCTING FLUCTUATIONS ABOVE \protect{$T_{c}$} AND COHERENCE OF 
THE ELECTRONIC SYSTEM}

All of the properties observed by Raman, ARPES, IR, and 
tunneling spectroscopies 
reveal an increase in the coherence of the electronic system while 
the underdoped cuprates are cooled toward $T_{c}$. 
The rearrangement of the Raman spectra with formation of the 75~meV mode 
is an indication that a partially coherent state forms above $T_{c}$ out 
of the incoherent QPs in the vicinity of $\{{\bf k}_{an}\}$ points. 
This long-lived state might consist of bound states of doped holes (preformed 
Cooper pairs) or of more complex many-body objects. 
Light scattering may break up the bound state into unbound  
QPs by a process similar to 2-M scattering.   
We propose that this process enhanced by a final state resonance is the 
origin of the 600~cm$^{-1}$ $B_{1g}$ Raman peak   
and suggest that it should be present for underdoped materials 
with AF correlations sufficient to exhibit underdamped 2-M excitations. 
The AF correlations that cover a few lattice spacings should 
last at least for $10^{-13}$~s, the lifetime of the coherent state. 

Occupancy of the coherent state lowers the density of the incoherent 
single holes,  
leading to reduction of the low-frequency inverse lifetime, 
suppression of the incoherent low-frequency density of states, and formation 
of the pseudogap.  
We note that because of different couplings to light 
and also because of the very different 
lifetimes of single holes and the bound state, 
there is no conservation of spectral weight for Raman scattering:  
the low-frequency intensity reduction is stronger 
than the increase of the intensity of the 600~cm$^{-1}$ mode. 
As seen from Fig.~3, the low-temperature reduction of the 
scattering rate, as also shown by optical data \cite{Puchkov96},  
occurs only below the 600~cm$^{-1}$ energy. 
Cooling increases the occupancy of the coherent state and 
enhances the pseudogap until phase ordering at $T_{c}$ produces a 
global SC state.   
Simultaneously, due to the phasing effect, the Raman spectra 
acquire additional 600~cm$^{-1}$ peak intensity.  

For the very underdoped sample ($T_{c} < 10$~K), only a weak pseudogap  
develops, and the integrated 600~cm$^{-1}$ peak intensity exhibits a 
weak enhancement (see Fig.~4).  
Both the pseudogap and the peak enhancement are stronger for the $T_{c} = 
65$~K sample. 
The sample with higher doping ($T_{c} \geq 83$~K) exhibits an effective 
condensation into the SC state with formation of a 
$2\Delta$-like feature in the Raman spectra out of the 600~cm$^{-1}$ peak. 
We note also that for all underdoped samples, only a partial 
condensation of single holes in the SC state occurs even at the 
lowest temperature. 

The coherent excitation near 600~cm$^{-1}$ above $T_{c}$ weakens with doping. 
Instead, a strong coherent $2\Delta$-peak, much stronger than the 600~cm$^{-1}$ 
peak for the underdoped materials, develops below $T_{c}$. 
For the overdoped sample, the $2\Delta$-peak shape, including its 
low-frequency tail, may be reproduced 
by Fermi-liquid based Raman scattering theory for $d-$wave SC with a large 
FS that predicts a cubic power law for the low-frequency 
tail of the $B_{1g}$ scattering intensity \cite{Devereaux}. 
This demonstrates that a large FS is restored for materials 
above the optimal doping and that at the higher dopings and 
low temperatures the incoherent scattering is greatly reduced up 
to about 400~cm$^{-1}$. 

\section{CONCLUSIONS}

Our analysis of a broad range of Raman data for cuprates at 
different dopings implies that for the underdoped superconductors:  

(i) Short-range AF order over a few lattice spacings 
persists upon doping from the AF insulator. 

(ii) Single quasi-particle excitations for ${\bf k}$ in the vicinity 
of $\{{\bf k}_{an}\}$ points are incoherent. 

(iii) In the presence of short range AF order,  
doped holes form long lived coherent state.

(iv) Binding of holes in the coherent state reduces the incoherent 
low-frequency scattering rate and a pseudogap forms in the spectra.   

(v) The occupancy of the coherent state increases with cooling 
until phase ordering at $T_{c}$ produces a global SC state. 

For overdoped superconductors, short range magnetic excitations become 
overdamped, no coherent QP excitations are observed above $T_{c}$, 
a strong coherent $2\Delta$-peak corresponding to the 
excitations across the superconducting gap develops below $T_{c}$.

\bigskip 

We benefit by discussions with  
D.N.\ Basov, A.V.\ Chubukov, V.J.\ Emery, S.A.\ Kivelson, P.\ Lee, 
D.K.\ Morr, D.\ Pines, P.M.\ Platzman and Y.J.\ Uemura. 
We are grateful to B.\ Dabrowski, W.C.\ Lee, and B.W.\ Veal for 
samples. 
The work was supported by NSF grant DMR 93-20892 (GB, MVK), 
NSF cooperative agreement DMR 91-20000 through the STCS (GB, MVK, 
PG, DH), ONR (CK), and DOE-BES contract W-31-109-ENG-38 (PG, DH).

\end{document}